\documentclass[prb,prb,reprint,superscriptaddress,aps,showpacs,showkeys,twocolumn]{revtex4-1}

\pdfoutput=1

\usepackage[utf8]{inputenc}
\usepackage[T1]{fontenc}
\usepackage{color}
\usepackage{graphicx}
\usepackage{units}

\begin{document}

\title{Strain controlled oxygen vacancy formation and ordering in CaMnO$_3$}
\date{\today}
\author{Ulrich Aschauer}
\affiliation{Materials Theory, ETH Zurich, Wolfgang-Pauli-Strasse 27, CH-8093 Z\"urich, Switzerland}
\author{Reto Pfenninger}
\affiliation{Materials Theory, ETH Zurich, Wolfgang-Pauli-Strasse 27, CH-8093 Z\"urich, Switzerland}
\author{Sverre M. Selbach}
\affiliation{Materials Theory, ETH Zurich, Wolfgang-Pauli-Strasse 27, CH-8093 Z\"urich, Switzerland}
\affiliation{Department of Materials Science and Engineering, Norwegian University of Science and Technology, NO-7491 Trondheim, Norway}
\author{Tor Grande}
\affiliation{Department of Materials Science and Engineering, Norwegian University of Science and Technology, NO-7491 Trondheim, Norway}
\author{Nicola A. Spaldin}
\affiliation{Materials Theory, ETH Zurich, Wolfgang-Pauli-Strasse 27, CH-8093 Z\"urich, Switzerland}

\begin{abstract}
We use first-principles calculations to investigate the stability of bi-axially strained \textit{Pnma} perovskite CaMnO$_3$ towards the formation of oxygen vacancies. Our motivation is provided by promising indications that novel material properties can be engineered by application of strain through coherent heteroepitaxy in thin films. While it is usually assumed that such epitaxial strain is accommodated primarily by changes in intrinsic lattice constants, point defect formation is also a likely strain relaxation mechanism. This is particularly true at the large strain magnitudes ($>$4\%) which first-principles calculations often suggest are required to induce new functionalities. We find a strong dependence of oxygen vacancy defect formation energy on strain, with tensile strain lowering the formation energy consistent with the increasing molar volume with increasing oxygen deficiency. In addition, we find that strain differentiates the formation energy for different lattice sites, suggesting its use as a route to engineering vacancy ordering in epitaxial thin films.
\end{abstract}

\maketitle

\section{Introduction} 

Materials of the ABO$_3$ perovskite family and derived structures exhibit a large variety of functional properties which are of fundamental and technological interest. Intrinsic and extrinsic point defects, such as oxygen vacancies or cation non-stoichiometry always exist in these materials and are in some cases desirable or even crucial for their properties. For example oxygen vacancies enable ionic conductivity in high temperature electrochemial devices such as oxygen sensors, SOFC cathodes and mixed conducting membranes \cite{Bouwmeester:2003im,Fleig:2003dq}. Deviations from ideal stoichiometry and charge balance are essential for introducing the carriers required for conventional- or super-conductivity, or carrier-mediated ferromagnetic ordering. The arrangement of defects can also be important, with reports of \textit{ordered} oxygen vacancies inducing high-temperature superconductivity \cite{Ourmazd:1987tx,Werder:1988wu} and multiferroicity \cite{Wei:2012if}. The electronic structure of interfaces in oxide heterostructures such as the formation of two-dimensional electron gases (2DEG)\cite{Mannhart:2010ha}, possibly relies on cation interdiffusion and will thus be critically affected by the concentration of cation vacancies required for cation migration. Point defects can also be detrimental, however, particularly if insulating behavior is desired, such as in ferroelectric or capacitor applications, or in achieving intrinsic antiferromagnetic ordering through superexchange.

In bulk perovskite oxides the presence of oxygen vacancies is well established to result in an expansion of the crystal lattice due to underbonding caused by the two extra electrons from the missing oxygen located in non-bonding orbitals \cite{Adler:2001ww,Ullmann:2001uc}. This phenomenon, known as chemical expansion, is most pronounced for perovskites containing transition metal ions that can adopt a variety of valence states and hence can readily accommodate the associated change in formal charge \cite{Marrocchelli:2012im}. Prominent examples are acceptor doped LaMnO$_3$ and donor doped SrTiO$_3$ \cite{Grande:2012ib, Moos:1997tu}, where significant volume expansions are observed as a function of the oxygen partial pressure.

\begin{figure}
\centering
\def\svgwidth{\columnwidth}
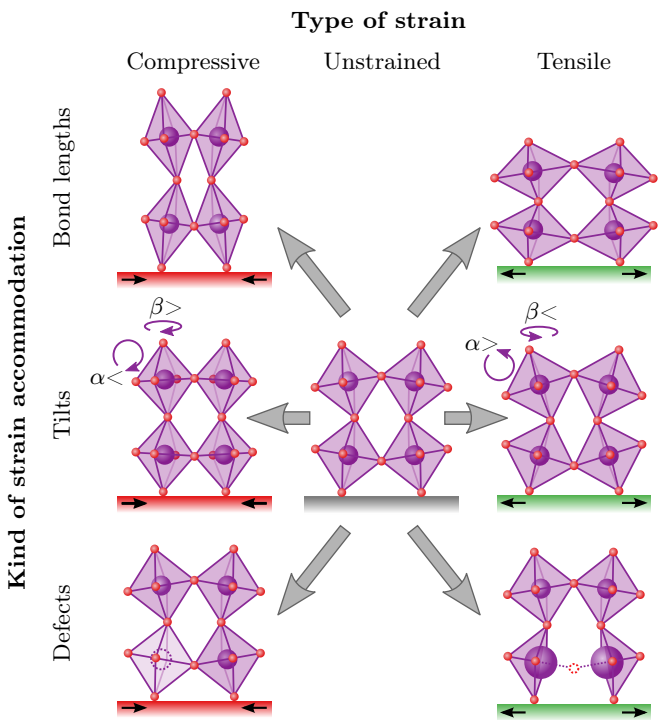
\caption{\label{fig:relaxation}Schematic illustration of possible mechanisms for epitaxial strain accommodation in perovskites. Only B-site octahedra are shown, A-site cations being omitted for clarity. Top row: accommodation by bond-length changes, center row: accommodation by changes in octahedral tilt ($\alpha$) and rotation ($\beta$) magnitudes, bottom row: accommodation by point defect formation. In the bottom right we illustrate the chemical expansion associated with oxygen vacancy formation, which is the subject of this work; the bottom left shows the ``chemical contraction'' caused by cation vacancy formation with concomitant oxidation of cations \cite{Grande:2012ib}.} 
\end{figure}

It has been demonstrated during the last decade that functional behavior in perovskites can be enabled or enhanced when they are prepared as thin films. For example, ferroelectricity has been reported in thin films of SrTiO$_3$, which is a quantum paraelectric in its bulk form, and enhanced polarization has been observed in thin films of ferroelectric BaTiO$_3$ \cite{Haeni:2004gj,Choi:2004it}. Reports of coupling magnetic and ferroelectric order parameters to yield thin-film magnetoelectric multiferroics \cite{Lee:2010kz,Gunter:2012ki} have spiked tremendous interest. The improved behavior is usually attributed to the bi-axial epitaxial strain introduced when the film is grown coherently on a substrate with a different lattice constant. In turn, it is assumed that the change in in-plane lattice constant associated with epitaxial strain is accommodated through structural distortions such as changing the internal bond lengths (top row of Fig. \ref{fig:relaxation}), or changing the angle or pattern of rotations and tilts of the oxygen octahedra (center row of Fig. \ref{fig:relaxation}) \cite{Megaw:1975ci,Thomas:1996ig,Rondinelli:2011jk}. Indeed first-principles electronic-structure calculations within this approximation have proven successful in predicting and explaining many of the observed novel behaviors \cite{Fennie:2006jp,Bhattacharjee:2009eb,Lee:2010kz}; for a review see Ref. \cite{Rondinelli:2011jk}.

In addition to changes in internal bond lengths or octahedral rotations, another possible effect of the change in lattice parameter forced by coherent epitaxy is a change in the defect profile. Such a response is likely because both strain and point defects affect the molar volume and hence the Gibbs free energy resulting in strain dependent defect formation energies. The increase in volume associated with biaxial tensile strain is likely to increase the concentration of oxygen vacancies (Fig. \ref{fig:relaxation}, lower right panel). In the opposite direction, compressive strain should increase the concentration of cation vacancies (Fig. \ref{fig:relaxation}, lower left panel) due to the concomitant charge balancing oxidation of cations with smaller ionic radii at higher oxidation states. Coherent epitaxial strain could thus be used as a parameter to control point defect populations. Conversely, the spontaneous formation of oxygen vacancies could act as a route of chemistry mediated strain relaxation \cite{Grande:2012ib,Selbach:2011bv} and allow large lattice mismatches to be accommodated in heterostructures. We explore these possibilities here.

\begin{figure}
\centering
\def\svgwidth{0.8 \columnwidth}
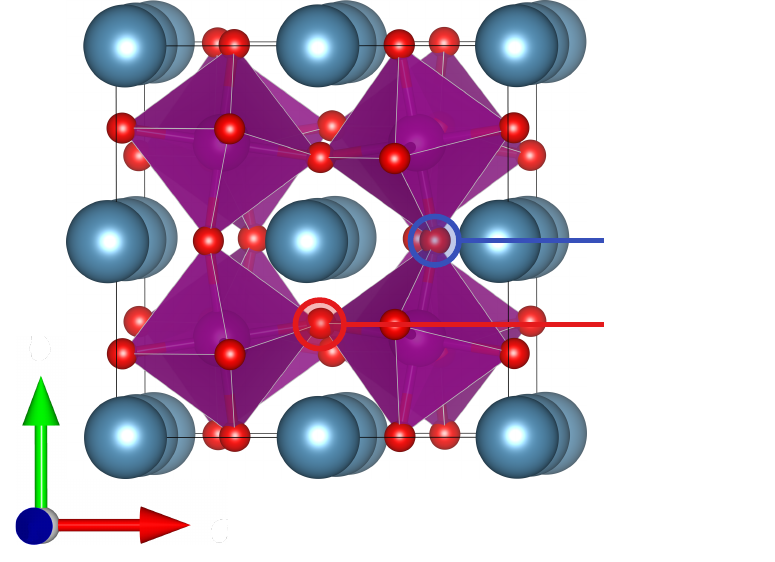
\caption{\label{fig:structure}The \textit{Pnma} crystal structure of bulk, unstrained CaMnO$_3$. The two inequivalent V$_\mathrm{O}$ positions, which are adjacent to Mn ions in the strained $ac$-plane (IP) and perpendicular to it (OP) are indicated. Color code: Ca: blue, Mn: purple, O: red.}
\end{figure}

We choose the perovskite CaMnO$_3$ as our model compound for analyzing the strain dependence of the oxygen vacancy (V$_\mathrm{O}$) formation reaction. CaMnO$_3$ forms in the widely adopted \textit{Pnma} modification of the ideal perovskite structure (Fig. \ref{fig:structure}), and this tilt pattern has been shown to remain stable throughout the entire range of likely experimentally accessible strains \cite{Bhattacharjee:2009eb,Gunter:2012ki,Bousquet:2011tf}. The oxygen vacancy has a charge of +2 relative to the perfect lattice, usually written as V$^{\bullet\bullet}_\mathrm{O}$ in Kr\"oger-Vink notation, but here V$_\mathrm{O}$ is used for simplicity. This relative charge of V$_\mathrm{O}$ has to be compensated by point defects with opposite charge to retain charge neutrality. The compensating point defect is Mn$^{3+}$ on Mn$^{4+}$ sites with a relative charge of -1, Mn$^{'}_{\mathrm{Mn}}$ in Kr\"oger-Vink notation. The chemical reaction that we study can be written as
\begin{equation}
\mathrm{2Mn^{4+} + O^{2-} \rightarrow 2Mn^{3+} + V_O + \nicefrac{1}{2} O_2 (g)} \label{eq:rxn}
\end{equation}
and has an experimental enthalpy of 1.89$\pm 0.04$ eV in bulk CaMnO$_3$ \cite{Rormark:2001kb}. Since high-spin Mn$^{3+}$ (0.645 Å) is larger than Mn$^{4+}$ (0.53 Å) \cite{Shannon:1976vx}, it is clear that reaction (\ref{eq:rxn}) should be accompanied by a considerable volume expansion, which indeed is supported by experiments \cite{Chen:2005cc,Zeng:1999us}. Since volume increases (decreases) with tensile (compressive) strain, we expect the V$_\mathrm{O}$ formation energy to be sensitive to strain. In addition to this volume effect, strain may also affect the V$_\mathrm{O}$ formation energy through its effect on the electronic energy levels. The electrons from the loss of O$_2$ which reduce Mn$^{4+}$ to Mn$^{3+}$ occupy the $e_g$ orbitals, the energies of which are strongly affected by the coordination geometry of the surrounding ions. 

Our main finding is that, as expected, tensile strain lowers the oxygen vacancy formation energy in CaMnO$_3$. This shows that the material will respond to tensile strain not only by by increasing its bond lengths and modifying its octahedral rotation angles, but also by reducing its oxygen stoichiometry by creation of oxygen vacancies. In addition, and perhaps unexpectedly, we find that strain modifies the oxygen vacancy formation energy of inequivalent oxygen sites markedly differently. This finding implies a possible strain control of oxygen vacancy ordering, which could enable new functionalities \cite{Borisevich:2012ku,Kalinin:2012fs}.

\section{Computational Details}

Density functional theory (DFT) calculations were performed using the VASP code \cite{Kresse:1993ty,Kresse:1994us,Kresse:1996vk,Kresse:1996vf}, with the PBEsol functional \cite{Perdew:2008fa} and wave functions expanded in plane waves up to a kinetic energy cutoff of 650 eV. The projector augmented wave (PAW) \cite{Blochl:1994uk,Kresse:1999wc} method was used to describe electron-core interactions with Ca(3s, 3p, 4s), Mn(3p, 3d, 4s) and O(2s, 2p) shells treated as valence electrons. The Hubbard U correction \cite{Anisimov:1991wt} was applied to Mn 3d electrons, using a U value of 3.0 eV, which was found to reproduce the experimentally determined density of states \cite{Jung:1997wq} and was recently shown to yield the correct stability of the different magnetic phases of CaMnO$_3$ \cite{Hong:2012dc} as well as the redox energetics of binary manganese oxides \cite{Wang:2006kn}. We used the ground-state G-type antiferromagnetic order in conjunction with a $2 \times 2 \times 2$ supercell of the 5 atom perovskite unit-cell. Reciprocal space integration was performed on a $4 \times 4 \times 4$ $\Gamma$-centered mesh. Structures were set up within the desired \textit{Pnma} space-group and subsequently relaxed until forces on the ions converged to below 10$^{-5}$ eV/\AA. In our calculations we strained CaMnO$_3$ in the plane of the equal $a$ and $c$ axes, while relaxing the $b$-axis length and all internal coordinates. During calculations containing V$_\mathrm{O}$ the lattice parameters were kept fixed respectively at the strained $a=c$ and relaxed $b$ lattice parameter, while relaxing only internal coordinates.

\section{Results \& Discussion}

\subsection {Stoichiometric CaMnO$_3$}

\begin{figure}
\centering
\def\svgwidth{\columnwidth}
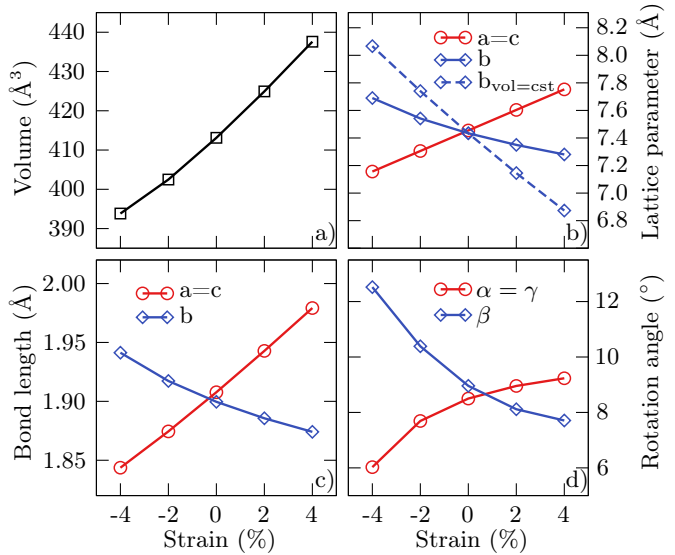
\caption{\label{fig:volume} Evolution of structural properties with biaxial strain in stoichiometric CaMnO$_3$: a) relaxed 8-formula-unit supercell volume (a material with an ideal Poisson ratio would have a constant volume with strain), b) lattice parameters (the dashed line shows the b-axis parameter for a material with ideal poisson ratio), c) Mn-O bond lengths and d) octahedral rotation angles (defined as angles between cartesian axes and Mn-O bonds).}
\end{figure}

We first look at strain induced changes in the stoichiometric CaMnO$_3$ structure, characterised by the volume, lattice parameters, bond lengths and octahedral rotation angles shown in Fig. \ref{fig:volume}. We notice first in Fig. \ref{fig:volume}(a), that strain leads to a net volume change in CaMnO$_3$, meaning that the material does not have a perfect Poisson ratio. This is also reflected in the lattice parameters shown in Fig. \ref{fig:volume}(b). While the $a$ and $c$ lattice parameters change linearly as a function of the imposed strain, the $b$ lattice parameter expands more strongly in the compressive range than it shrinks in the tensile range. It's rate of change is however about a factor two smaller than required for constant volume, leading to the observed volume decrease with increasing compression. 

As schematically shown in in Fig. \ref{fig:relaxation}, in the stoichiometric structure, these changes in lattice parameter are accommodated by changes in bond-lengths and octahedral rotation angles. Indeed in Fig. \ref{fig:volume}(c) and (d) we see that both the bond lengths and tilt angles change as a function of bi-axial strain. The bond lengths are however relatively insensitive to strain (an order of magnitude smaller change than the lattice parameters, see Fig. \ref{fig:volume}(b)), meaning that for stoichiometric \textit{Pnma} CaMnO$_3$ bi-axial strain is primarily accommodated by altering the octahedral tilt angles, Fig. \ref{fig:volume}(d). As expected, the octahedral tilt angles increase with compression of the respective axis, and decrease when the axis becomes longer.

\subsection{Tensile strain}

\begin{figure}
\centering
\def\svgwidth{\columnwidth}
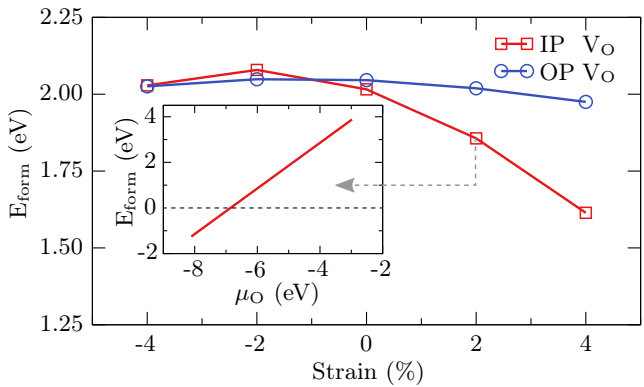
\caption{\label{fig:formation}V$_\mathrm{O}$ formation energies for the two inequivalent V$_\mathrm{O}$ positions in the \textit{Pnma} tilted CaMnO$_3$ structure at an oxygen chemical potential of -5 eV, corresponding to typical growth conditions under air. The inset shows the formation energy of the IP V$_\mathrm{O}$ at 2\% tensile strain as a function of the oxygen chemical potential. The OP V$_\mathrm{O}$ very closely tracks this curve and is omitted for clarity.}
\end{figure}

In the orthorhombic CaMnO$_3$ structure there are two inequivalent oxygen positions as shown in Fig. \ref{fig:structure}. We label these IP and OP, corresponding to the broken Mn-O bonds lying in- or out-of the plane of epitaxial strain respectively, but note that the sites are already inequivalent in the unstrained \textit{Pnma} structure. In Fig. \ref{fig:formation} we show our calculated formation energies for these two oxygen vacancies as a function of the applied strain at an oxygen chemical potential of -5 eV, corresponding to typical growth conditions under air. In the inset in Fig. \ref{fig:formation} we show the V$_\mathrm{O}$ formation energy in the 2\% tensile strained structure as a function of the oxygen chemical potential throughout the whole range of CaMnO$_3$ thermodynamic stability evaluated by computing the Ca-Mn-O ternary phase diagram \cite{Luo:2009gg}. As is well known, the oxygen chemical potential strongly affects the absolute value of the formation energy and hence the equilibrium oxygen vacancy concentration.

\subsubsection{Change in stoichiometry}

As expected from the known chemical expansion caused by oxygen vacancies, we find that tensile strain lowers the formation energy for oxygen vacancies in CaMnO$_3$. The formation energy decreases considerably by $\sim$0.4eV corresponding to $\sim$20\% of the bulk value for formation of the IP vacancy at 4\% strain. 

In Fig. \ref{fig:implications}(a) we show the calculated equilibrium change in stoichiometry as a function of temperature for tensile strain values of 0 and 4\% and oxygen partial pressures corresponding to growth in air (0.21) and under nitrogen (10$^{-6}$). The absolute oxygen stoichiometry 3-$\delta$ was estimated from reaction (\ref{eq:rxn}), the equilibrium constant of which can be written as \cite{Bakken:2005dd}:
\begin{equation}
\mathrm{K = \frac{{\left[Mn^{3+}\right]}^2 \left[V_O\right] p_{O_2}^{\nicefrac{1}{2}}}{{\left[Mn^{4+}\right]}^2 \left[O^{2-}\right]}= exp\left(-\frac{\Delta G_f}{k_B T}\right)} \label{eq:K1}
\end{equation}
Here G$_\mathrm{f}$ is the Gibbs energy of reaction (\ref{eq:rxn}). From reaction (\ref{eq:rxn}) and charge neutrality we get $\mathrm{\left[V_O\right]} = \delta$, $\mathrm{\left[Mn^{3+}\right]} = 2\delta$, $\mathrm{\left[Mn^{4+}\right]} = 1 - 2\delta$ and $\mathrm{\left[O^{2-}\right]} = 3 - \delta$. Equation (\ref{eq:K1}) was solved for $\delta$ using $\Delta\mathrm{G_f} = \Delta\mathrm{H_f - T \Delta S}$ where $\Delta \mathrm{H_f}$ is the enthalpy of oxygen vacancy formation and $\Delta \mathrm{S}$ the entropy of reaction (\ref{eq:rxn}), respectively, where the latter is 95 JK$^{-1}$mol$^{-1}$ taken from \cite{Bakken:2005dd}.

At a typical growth temperature of 1000 K (indicated by the vertical dashed line in Fig. \ref{fig:implications}), tensile strain of 4\% reduces the equilibrium oxygen stoichiometry from 2.995 to 2.978 in air and from 2.964 to 2.859 in N$_2$ atmosphere (p$_\mathrm{O_2}$ = 10$^{-6}$ atm.), as illustrated in Fig. \ref{fig:implications}. The oxygen stoichiometry of the supercell used in our calculations is 3-$\delta$ = 2.875 for comparison. A reduction of 0.4 eV of the V$_\mathrm{O}$ formation energy has an impact on $\delta$ equivalent to raising the temperature 150-200 K, and will drastically alter the equilibrium oxygen content. 

\subsubsection{Vacancy ordering}

While both vacancy types lower their formation energy, we find that the energy lowering is stronger for the IP vacancy than for the OP, consistent with the fact that tensile strain expands the in-plane lattice parameter and leads to a slight contraction of the out-of plane dimensions (see Fig. \ref{fig:volume} (b)), however still with a net increase in volume as shown in Fig. \ref{fig:volume}(a). This suggests that, in addition to tuning the oxygen vacancy formation energy, tensile bi-axial strain could be used to influence vacancy site preference and thus induce vacancy ordering. 

\begin{figure}
\centering
\def\svgwidth{\columnwidth}
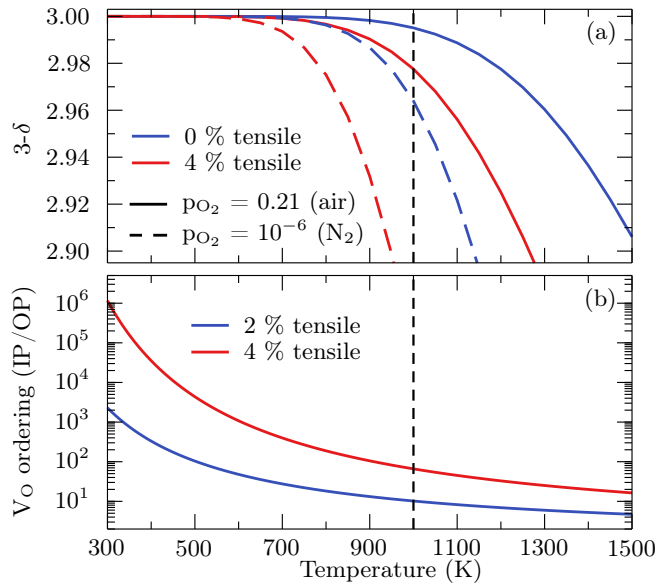
\caption{\label{fig:implications}a) Modelled oxygen stoichiometry of CaMnO$_{3-\delta}$. The 3-$\delta$ curves are calculated from equation (\ref{eq:K1}), see text for explanation. b) Temperature dependent V$_\mathrm{O}$ ordering expressed as the ratio of IP/OP (Fig. \ref{fig:structure}) for +2\% and +4\% tensile strain (Fig. \ref{fig:formation}). The typical growth temperature of 1000K is indicated by the vertical dashed line.}
\end{figure}

In Fig. \ref{fig:implications}(b) we show our calculated equilibrium ratio of IP to OP V$_\mathrm{O}$ as a function of temperature for tensile strain values of 2 and 4\%. The ratio of IP to OP vacancies under tensile strain was calculated from the shift in formation energies shown in Fig. \ref{fig:formation}. Under +4\% tensile strain, for example, the IP V$_\mathrm{O}$ formation energy is 0.36 eV lower than that of the OP V$_\mathrm{O}$. At the typical growth temperature of 1000K \cite{Zheng:2010ku} (indicated by the vertical dashed line in Fig. \ref{fig:implications}) this energy difference corresponds to an equilibrium ratio of IP to OP V$_\mathrm{O}$ = $\sim$73. However, oxygen vacancies can remain mobile to much lower temperatures \cite{Bouwmeester:2003im,Fleig:2003dq,Chen:2013hp}. Equilibrium vacancy concentrations can thus be reached at significantly lower temperatures, especially in thin films where the diffusion distances are short \cite{Grande:2012ib}. At 300 K the equilibrium ratio of IP to OP vacancies is more than 10$^6$, (see Fig. \ref{fig:implications}(b) demonstrating the possibility of engineering oxygen vacancy ordering by epitaxial strain. Oxygen vacancy ordering induced by epitaxial strain modification of the vacancy formation energy can in principle be applied to \textit{any} oxide with non-equivalent oxygen sites. This can pave the way towards new interface functionalities arising from orbital ordering (Fig. \ref{fig:pdos} (d)), localized electric dipoles/fields \cite{Borisevich:2012ku} or enhanced interface electronic, ionic or mixed conductivity.

\subsection{Compressive strain}

In spite of the fact that the volume of the cell decreases monotonically from tensile to compressive strain as shown in Fig. \ref{fig:volume}(a), we find a negligible change in the energy of formation of either vacancy on compressive strain. While our findings for tensile strain are consistent with the simple arguments based on changes in Mn ion valence and volume presented in the introduction, the compressive-strain response is fundamentally different. Next we analyze the changes that strain induces in the electronic structure to understand the origin of this behavior.

\begin{figure}
\centering
\def\svgwidth{\columnwidth}
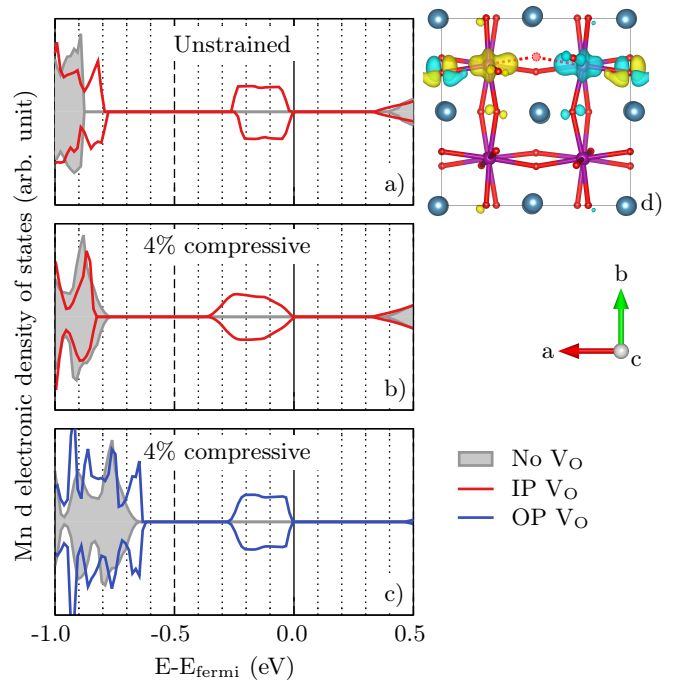
\caption{\label{fig:pdos}Mn \textit{d} partial density of states shown for (a) the unstrained IP V$_\mathrm{O}$ case, (b) the IP V$_\mathrm{O}$ at 4\% compressive strain and (c) the OP V$_\mathrm{O}$ at 4\% compressive strain. The DOS of the stoichiometric material is shown in grey for the corresponding strain. The energy scales of the stoichiometric DOS have been aligned with respect to the defective DOS at a Mn \textit{p} core-state about 50 eV below the Fermi energy. The origin in energy is set at the Fermi energy of the defective cell. Panel (d) shows the charge density associated with the occupied $e_g$ states upon IP V$_\mathrm{O}$ formation. The dashed red lines and circle indicate the bonds and position of the missing O atom. Yellow and light blue isosurfaces  at $8\cdot 10^{-3}$ e\AA$^{-3}$ show the up and down spin density respectively. Color code: Ca: blue, Mn: purple, O: red.}
\end{figure}

In Fig. \ref{fig:pdos} (a) we show our calculated electronic density of states (DOS) for unstrained CaMnO$_3$ with one out of 24 O atoms missing. We see that IP V$_\mathrm{O}$ formation produces an extra peak within the band-gap of the density of states. The OP V$_\mathrm{O}$ produces a peak at nearly the same position and has hence been omitted for clarity. The IP V$_\mathrm{O}$ peak accommodates the two charge compensating electrons, which are primarily located in the formerly unoccupied $z^2$-type orbitals of the two Mn atoms adjacent to the oxygen vacancy, as shown in the calculated charge density in Fig. \ref{fig:pdos} (d). Localisation of these electrons is a result of the removal of the anti-bonding $\sigma^*$ overlap between the Mn $e_g$ and the O p orbitals along the axis of the broken bond, which lowers the energies of these $e_g$ orbitals.

We also observe that both vacancies will cause an upward shift of about $\sim$0.1 eV of the valence band edge, which consists of Mn $t_{2g}$ orbitals. For the IP structure the increase stems from the xy and xz orbitals, whereas in the OP structure the xy and yz orbitals move up in energy. This reflects the loss in Mn-O $\pi$ bonding overlap with the bond broken by IP V$_\mathrm{O}$ formation lying along the a (x) axis, whereas that broken by OP V$_\mathrm{O}$ lies along the b (y) axis.

In Fig. \ref{fig:pdos} (b) and (c) we look at the changes in band edge and gap state in the DOS of the IP and OP structure compared to the stoichiometric case. We can see that the valence band edge for the IP V$_\mathrm{O}$ is lowered by $\sim$ 0.05 eV, whereas for the OP V$_\mathrm{O}$ is raised by an equal amount. We can again rationalize these effects in terms of structural changes. In the stoichiometric and IP case, the valence band-edge consists primarily of xz orbitals lying in the compressed plane. Creation of an V$_\mathrm{O}$ in this plane can to some extent counteract the increase in anti-bonding overlap induced by compression and hence lead to a slight net lowering of the band edge energy in the IP case. This is not the case for the OP V$_\mathrm{O}$, for which moreover the xy and yz orbitals increase in energy due to loss of Mn-O bonding overlap, resulting in the observed net upwards shift. 

The opposite trend is observed for the gap state, which is lowered in energy by $\sim$ 0.1 eV for the OP case, whereas in the IP case we observe a broadening with the lower edge dropping by $\sim$ 0.05 eV and the upper edge remaining more or less constant. The downwards shift of the OP gap state is due to decreased $\sigma^*$ overlap since the broken Mn-O bonds lie along the stretched b axis.  Conversely the broadening for the IP case stems from an increase of this overlap due to compression of the Mn-O bonds lying in the compressively strained plane.

In the compressive regime we thus have counterbalancing effects, leading to very similar formation energies for the two vacancies, which are moreover almost independent of strain. 

\begin{figure}
\centering
\def\svgwidth{\columnwidth}
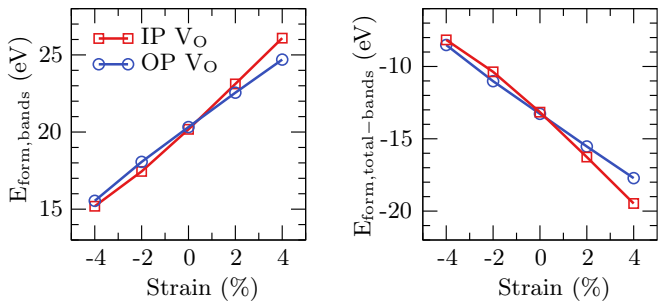
\caption{\label{fig:decompose}The total formation energy shown in Fig. \ref{fig:formation} decomposed into contributions from a) the band energy and b) the remaining interactions.}
\end{figure}

The above results indicate that crystal field effects dominate changes in V$_\mathrm{O}$ formation energies under compression. This can be further assessed by a decomposition of the V$_\mathrm{O}$ formation energy into its components. In Fig. \ref{fig:decompose} we show that in the tensile regime the band energy related part (Fig. \ref{fig:decompose}a) alone would predict the OP V$_\mathrm{O}$ to be lower in energy. In order to recover the observation in Fig. \ref{fig:formation} that the IP V$_\mathrm{O}$ is more stable, the remaining - mainly electrostatic - interactions (Fig. \ref{fig:decompose}b) have to dominate in the tensile range. The marked difference between IP and OP under tension is a result of the drastic reduction in repulsive electron-electron interaction due to stretching of the broken bond axis. Conversely in the compressive range in Fig. \ref{fig:decompose}, we can see that the observed behaviour of the total formation energy (Fig. \ref{fig:formation}) stems mainly from changes in the band energy due to crystal field effects.

In our calculations relaxation of the internal coordinates allowed both octahedral rotation angles and bond lengths to vary as a function of strain. As mentioned earlier (Fig. \ref{fig:volume}), changes in octahedral rotation angles account for a large part of strain accommodation in the \textit{Pnma} structure. Therefore crystal field-effects due to bond-length changes should be rather weak. In order to asses the limit of strong crystal-field effects, when no changes in tilt pattern exist, we have performed the same analysis on the ideal cubic structure. Under tensile strain both vacancies show a similar although less marked behavior than in the \textit{Pnma} structure. In the compressive range however we find the OP V$_\mathrm{O}$ formation energy to continuously increase whereas the IP V$_\mathrm{O}$ shows a marked decrease. This is the result of a pronounced broadening of the $e_g$ gap states in the cubic structure as the strain is accommodated by changes in bond-length (overlap) only. These observations confirm again that under compression crystal-field effects dominate whereas electrostatics are at the origin of the observed behaviour under tensile strain.

\section{Summary \& Conclusions}

In the present work we have investigated how the oxygen vacancy formation energy is modified by bi-axial strain from 4\% compressive to 4\% tensile strain. Our DFT calculations show that oxygen vacancy energetics are strongly modified by strain, particularly by tensile strain which was found to favor the formation of oxygen vacancies. Compared to the unstrained material, the formation energy of oxygen vacancies is reduced by up to 0.4 eV under 4\% tensile strain, significantly reducing the equilibrium oxygen content in films under tensile strain. The origin of this reduction is mainly electrostatic, as tensile strain reduces the electron-electron repulsion along the broken Mn-O-Mn bond axis. It thus follows that CaMnO$_3$ can accommodate strain by increasing its concentration of oxygen vacancies \cite{Grande:2012ib,Selbach:2011bv,Venkatesan:2011fz}.

Compressive strain on the other hand was found to affect Mn d levels in a more subtle way with nearly no net effect on oxygen vacancy formation energies. Although not explicitly treated in this work, we expect the cation vacancy concentration in oxygen hyperstoichiometric perovskites, like LaMnO$_{3+\delta}$, to be correspondingly sensitive to compressive epitaxial strain \cite{Grande:2012ib}. In such materials the hyperstoichiometry is a result not of oxygen excess but cation deficiency.

This stabilization of oxygen vacancies by tensile strain is the reverse effect of the well established chemical expansion phenomenon where oxygen vacancies induce a unit cell expansion \cite{Adler:2001ww,Ullmann:2001uc}. Our findings have profound implications for the properties, growth and thermal annealing of oxide thin films under tensile strain. Epitaxial strain, like temperature and partial pressure of oxygen (p$_\mathrm{O_2}$), is a thermodynamic variable affecting the equilibrium oxygen stoichiometry.

The formation of oxygen vacancies must be charge compensated, either by reducing the formal oxidation state of cations (e.g. Mn$^{4+}$ to Mn$^{3+}$ in the present work) or more unfavorably by donating electrons into the high energy conduction band (e.g. LaAlO$_3$ \cite{Luo:2009gg}). The formation of oxygen vacancies is therefore likely to be enhanced by tensile strain in \textit{any} oxide with transition metal cations with multiple oxidation states such as Mn, Fe, Co, Ni or even Cu. The coupling between strain and oxygen vacancy formation energy is moreover expected to be most pronounced in materials containing transition metal cations with localized $d$-electrons \cite{Jalili:2011hb,Donner:2011hl,Kushima:2010ex}.

Our calculations also show that tensile strain affects the formation energy of the two kinds of non-equivalent oxygen vacancies in CaMnO$_3$ with \textit{Pnma} perovskite structure in a markedly different way. Under 4\% tensile strain the formation energy of oxygen vacancies on the in-plane site is 0.36 eV lower than on the out-of-plane site. Depending on the temperature this large energy difference may result in orders of magnitude differences in vacancy concentrations on the two sites and paves the way for tensile-strain-induced oxygen vacancy ordering. This possibility of engineering epitaxial heterostructures with oxygen vacancy ordering should apply to \textit{any} oxide with non-equivalent oxygen sites.

\bibliography{references.bib}

\end{document}